\documentstyle[12pt,epsf]{article}

\newcommand{\bda}{\begin{displaymath}\begin{array}{rl}}
\newcommand{\eda}{\end{array}\end{displaymath}}
\newcommand{\be}{\begin{equation}}
\newcommand{\ee}{\end{equation}}
\newcommand{\bea}{\begin{eqnarray}}
\newcommand{\eea}{\end{eqnarray}}
\newcommand{\bdm}{\begin{displaymath}}
\newcommand{\edm}{\end{displaymath}}
\newcommand{\no}{\nonumber \\}

\newcommand{\ubar}{\overline{\rule[0.42em]{0.4em}{0em}}\hspace{-0.5em}u}
\newcommand{\dbar}{\,\overline{\rule[0.7em]{0.4em}{0em}}\hspace{-0.6em}d}
\newcommand{\sbar}{\overline{\rule[0.45em]{0.4em}{0em}}\hspace{-0.5em}s}

\newcommand{\R}{{\scriptscriptstyle R}}

\newcommand{\al}{&\!\!\!\!}
\newcommand{\fs}{\; \; .}
\newcommand{\co}{\; \; ,}
\newcommand{\QD}{Q_{\hspace{-0.1em}D}}

\begin{document}

\begin{titlepage}

\begin{flushright}
CERN-TH/96-44\\
hep-ph/
\end{flushright}

\vspace{2cm}
\begin{center}
{\LARGE {\bf The ratios of the light quark masses}}
\\ \vspace{0.8cm}
H. Leutwyler\\Institut f\"{u}r theoretische Physik der Universit\"{a}t
Bern\\Sidlerstr. 5, CH-3012 Berne, Switzerland and\\ \rule{0cm}{0cm}\\
CERN, CH-1211 Geneva 23\\ Switzerland\\
\vspace{0.6cm}
\vspace{0.6cm}
{\bf Abstract} \\
\vspace{1.2em}
\parbox{28em}{The paper collects the various pieces of information
concerning the relative size of $m_u,\,m_d$ and $m_s$. A coherent picture
results, which constrains the mass ratios to a rather narrow range:
$m_u/m_d\!=\!0.553\pm0.043$, $m_s/m_d\!=\!18.9\pm 0.8$.}

\vspace{4cm}
\begin{flushleft}CERN-TH/96-44\\
February 1996\end{flushleft}
\rule{30em}{.02em}\\
{\footnotesize Work
supported in part by Schweizerischer Nationalfonds}
\end{center}
\end{titlepage}

1. I wish to show that recent results of chiral perturbation theory allow
a rather accurate determination of the relative size of $m_u,\,m_d$ and $m_s$.
The paper amounts to an update of 
earlier work \cite{Weinberg 1977}--\cite{HL Dallas Florida ITEP} based on the 
same method. Sum rules and numerical simulations of
QCD on a lattice represent alternative approaches with a broader scope --
they permit a determination not only of the ratios $m_u:m_d:m_s$, but also of
the individual quark masses, including the heavy
ones. The sum rule results for the ratios
are subject to comparatively large errors \cite{QCD SR,BPdR}.
Concerning the lattice technique, considerable progress has been
made \cite{lattice,Duncan Eichten Thacker}. It is difficult, however,
to properly account for the vacuum fluctuations generated by quarks with small
masses. Further progress with light
dynamical fermions is required before the numbers obtained for $m_u/m_d$ or
$m_s/m_d$ can be taken at face value.

2. The quark masses depend on the renormalization scheme.
Chiral perturbation theory treats the mass term of the light quarks,
$m_u\,\ubar u+m_d \,\dbar d+ m_s\, \sbar s$, as a perturbation \cite{Weinberg
Physica,GL SU(3)}.
It exploits the fact that, {\it for mass independent renormalization schemes},
the operators $\ubar u$, $\dbar d$ and
$\sbar s$ transform as members of the representation $(3,3^*)+(3^*,3)$.
Since all other operators with this transformation property are of higher
dimension, the normalization conventions for
$m_u$, $m_d$ and $m_s$ then differ only by a flavour-independent factor. 
The factor, in particular, also depends
on the renormalization scale, but in
the ratios $m_u/m_d$ and $m_s/m_d$, it drops out -- these represent
convention-independent pure numbers.

3. The leading order mass formulae for the Goldstone bosons follow from 
the relation of Gell-Mann, Oakes and Renner.
Disregarding the
electromagnetic interaction, they read $M_{\pi^+}^2=(m_u+m_d)B$,
$M_{K^+}^2=(m_u+m_s)B$, $M_{K^0}^2=(m_d+m_s)B$, where the constant of 
proportionality is determined by the quark condensate:
$B\!=\!|\langle0|\,\ubar u|0\rangle|/F_\pi^2$. Solving for the quark masses
and forming ratios, this constant drops out, so that $m_u/m_d$ and $m_s/m_d$
may be expressed in terms of ratios of meson masses. Current algebra also
shows that the mass difference between the $\pi^+$ and the $\pi^0$ is
almost exclusively due to the electromagnetic interaction -- the
contribution generated by $m_d\neq m_u$ is of order $(m_d-m_u)^2$ and
therefore tiny. Moreover, the Dashen theorem states that,
in the chiral limit, the electromagnetic
contributions to $M_{K^+}^2$ and to $M_{\pi^+}^2$ are the same, while the self
energies of $K^0$ and $\pi^0$ vanish. Using these relations to correct for the
electromagnetic self energies, the above lowest order mass formulae yield
\cite{Weinberg 1977}
\bea \label{r3}
\frac{m_u}{m_d} & =& \frac{M^2_{K^+}-  M^2_{K^0} + 2 M^2_{\pi^0} -
M^2_{\pi^+}} {M^2_{K^0} - M^2_{K^+} + M^2_{\pi^+}}= 0.55\co \\
\frac{m_s}{m_d} & = & \frac{M^2_{K^0} + M^2_{K^+} - M^2_{\pi^+}}
{M^2_{K^0} - M^2_{K^+} + M^2_{\pi^+}}= 20.1\nonumber\fs
\eea

4. These relations and the analysis given below are based on the
hypo\-thesis that the quark condensate is the leading order
parameter of the spontaneously broken symmetry. This  hypo\-thesis is 
questioned in refs.
\cite{Stern}, where a more general scenario is described,
referred
to as generalized chiral perturbation theory: 
Stern and co-workers investigate the possibility that the correction of
$O(m^2)$ in the expansion
$M_\pi^2\!=\!(m_u\!+\!m_d)B\!+\!O(m^2)$ 
is comparable with or even larger than the term that originates in the quark 
condensate.
Indeed, the available evidence does not exclude this
possibility, but a beautiful experimental proposal has been made
\cite{Nemenov}: $\pi^+\pi^-$ atoms decay into a pair of neutral pions, through
the strong transition $\pi^+\pi^-\!\rightarrow\!\pi^0\pi^0$. Because the
momentum transfer nearly vanishes, the decay rate is determined by the
combination $a_0\!-\!a_2$ of S-wave $\pi\pi$ scattering lengths. Since chiral
symmetry implies that Goldstone bosons of
zero energy do not interact, $a_0,a_2$ vanish in the limit
$m_u,m_d\!\rightarrow\!0$. The transition amplitude, therefore, directly
measures the symmetry breaking generated by $m_u,m_d$.
Standard chiral
perturbation theory yields very sharp predictions for $a_0,a_2$
\cite{scattering lengths}, while the
generalized scenario does not \cite{Stern two loops}. A measurement of the
lifetime of a $\pi^+\pi^-$ atom would thus
allow us to decide whether or not the quark condensate represents the
leading order parameter.

5. The contributions of first non-leading order were worked out in ref.
\cite{GL SU(3)}. As is turns out, the
correction in the
mass ratio $(M_{K^0}^2-M_{K^+}^2)/(M_K^2-M_\pi^2)$ is the same as the one in
$M_K^2/M_\pi^2$:
\bea
\label{r1}\frac{M_K^2}{M_\pi^2}\al=\al
\frac{\hat{m}+m_s}{m_u+m_d}\{1+\Delta_M+O(m^2)\}\co\\
\frac{M_{K^0}^2-M_{K^+}^2}{M_K^2-M_\pi^2}\al=\al\frac{m_d-m_u}{m_s-\hat{m}}
\{1+\Delta_M +O(m^2)\}\fs\nonumber\eea
The quantity $\Delta_M$ accounts for the breaking of SU(3) and is related to
the effective coupling constants $L_5$ and $L_8$:
\be
\label{DeltaM}\Delta_M=
\frac{8(M_K^2-M_\pi^2)}{F_\pi^2}\,(2L_8-L_5)+\chi\mbox{logs}\fs\ee
The term $\chi$logs stands for the logarithms characteristic of
chiral perturbation theory (for an explicit expression, see \cite{GL SU(3)}).
The above relations imply that the double ratio
\be\label{r2}
Q^2 = \frac{m^2_s - \hat{m}^2}{m^2_d - m^2_u} \ee
is given by a ratio of meson masses, up to corrections of second order,
\be\label{defQ}
Q^2 \equiv \frac{M^2_K}{M_\pi^2}\cdot \frac{ M^2_K - M^2_\pi}{M^2_{K^0} -
M^2_{K^+}}\,\{1+O(m^2)\}\fs \end{equation}
The result may be visualized by
plotting 
$m_s/m_d$ versus $m_u/m_d$ \cite{Kaplan Manohar}. The constraint then
takes the form of an ellipse,
\be\label{ellipse}
\left ( \frac{m_u}{m_d} \right)^2 + \,\frac{1}{Q^2} \left ( \frac{m_s}{m_d}
\right)^2 = 1\co
\end{equation}
with $Q$ as major semi-axis, the minor one being equal to 1 (for simplicity,
I have discarded the term
$\hat{m}^2/m_s^2$, which is numerically very small).

6. The meson masses occurring in the double ratio (\ref{defQ}) refer to pure
QCD. Correcting
for the electromagnetic self energies with the Dashen theorem, the quantity $Q$
becomes
\be\label{QD} \QD^{\;2}=
\frac{(M_{K^0}^2+M_{K^+}^2-M_{\pi^+}^2+M_{\pi^0}^2)
(M_{K^0}^2+M_{K^+}^2-M_{\pi^+}^2-M_{\pi^0}^2)}
{4\,M_{\pi^0}^2\,(M_{K^0}^2-M_{K^+}^2+M_{\pi^+}^2-M_{\pi^0}^2)}\fs\ee
Numerically, this yields $\QD=24.2$.
For this value of the semi-axis, the ellipse passes through the point
specified
by Weinberg's mass ratios, which correspond to $\Delta_M\!=\!0,\,Q\!=\!\QD$.

The Dashen theorem is subject to corrections from higher order terms in the
chiral expansion, which
are analysed in several
recent papers.
Donoghue, Holstein and Wyler \cite{DHW em self energies} estimate the
contributions arising from vector meson exchange and conclude
that these give rise to large corrections, increasing the
value $(M_{K^+}\!-\!M_{K^0})_{e.m.} \!=\! 1.3$ MeV predicted by Dashen to $2.3$
MeV. According to Baur and Urech \cite{Urech}, however, the model used is in
conflict with chiral symmetry:
although the perturbations due to vector meson exchange are enhanced
by a relatively small energy denominator, chiral symmetry prevents
them from being large. In view of this, it is puzzling that
Bijnens \cite{Bijnens}, who evaluates the self energies within the model of
Bardeen et al., finds an even larger effect,
$(M_{K^+}\!-\!M_{K^0})_{e.m.}\! \simeq\! 2.6\,\mbox{MeV}$.
The implications of the above estimates for the value of
$Q$ are illustrated on the r.h.s. of fig. 1.

Recently,
the electromagnetic self energies have been analysed within
lattice QCD \cite{Duncan Eichten Thacker}. The
result of this calculation, $(M_{K^+}\!-\!M_{K^0})_{e.m.} \!=\! 1.9$ MeV,
indicates that the corrections
to the Dashen theorem are indeed substantial, although not quite as large as
found in refs. \cite{DHW em self energies,Bijnens}.
The uncertainties of the lattice result are of the same type as those
occurring in direct determinations of the quark masses with this method. The
mass difference between $K^+$ and
$K^0$, however, is predominantly due to $m_d\!>\!m_u$, not to the e.m.
interaction. An error of 20\% in the self energy affects the value of
$Q$ by only about 3\%. The terms neglected when evaluating $Q^2$ with
the meson masses are of order $(M_K^2-M_\pi^2)^2/M_0^4$, where $M_0$ is
the mass scale relevant for the exchange of scalar or pseudoscalar
states, $M_0\!\simeq\! M_{a_0}\!\simeq\! M_{\eta'}$ \cite{HL 1990}. The
corresponding error in the result for $Q$ is also of the order of 3\% --
the uncertainties in
the value $Q\!=\!22.8$ that follows from the lattice result are significantly
smaller than those obtained for the quark masses with the same method.
\begin{figure}[t] \centering
\mbox{ \epsfbox{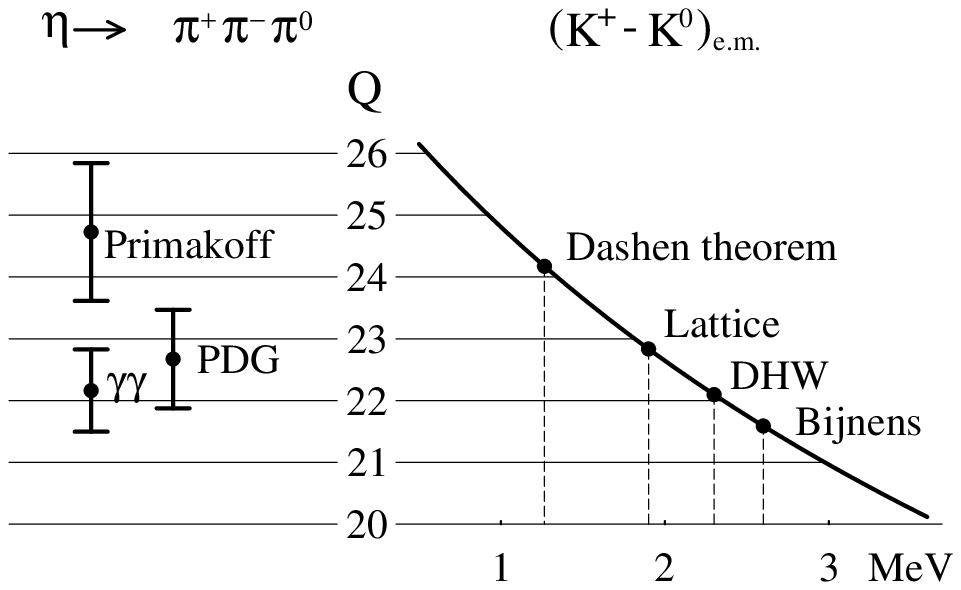} }
\parbox{13.7cm}{Figure 1:
The l.h.s. indicates the values of $Q$ corresponding to the various 
experimental
results for the rate
of the decay $\eta\!\rightarrow\!\pi^+\pi^-\pi^0$. The r.h.s. shows
the results for $Q$ obtained with four different theoretical estimates for
the electromagnetic self energies of the kaons.}\end{figure}

7. The isospin-violating decay $\eta \rightarrow 3\pi$ allows one to measure
the semi-axis in an entirely independent manner \cite{GL eta}. The transition
amplitude is much less sensitive to the
uncertainties associated with the electromagnetic interaction than the
$K^0\!-\!K^+$ mass difference: the e.m. contribution is
suppressed by chiral symmetry and is negligibly small \cite{Baur Kambor Wyler}.
The decay $\eta\rightarrow 3\pi$ thus represents a sensitive probe of the
symmetry breaking generated by $m_d-m_u$.
It is convenient to write the decay rate in the form
$\Gamma_{\eta \rightarrow \pi^+\pi^-\pi^0} \!=\!
\Gamma_0\,(\QD/Q)^4$, where $\QD$ is specified in eq. (\ref{QD}). As shown in
ref. \cite{GL eta}, chiral perturbation theory to one loop
yields a parameter-free prediction for the constant
$\Gamma_0$.
Updating the value of $F_\pi$, the numerical result reads
$\Gamma_0\!=\!168\pm50\,\mbox{eV}$.
Although the calculation includes
all corrections of
first non-leading order, the error bar is large. The problem originates in the
final state interaction, which strongly amplifies the transition probability
in part of the Dalitz plot. The one-loop calculation does account for this
phenomenon, but only to leading order in the low energy expansion.
The final state interaction is analysed more accurately in two recent papers
\cite{Kambor Wiesendanger Wyler,AL}, which exploit the fact that analyticity
and unitarity determine the
amplitude up to a few subtraction constants. For these, the corrections to the
current algebra predictions are small, because they are barely
affected by the final state interaction. Although the dispersive framework
used in the two papers differs, the results are nearly the same: while
Kambor, Wiesendanger and Wyler obtain
$\Gamma_0\!=\!209\pm20\,\mbox{eV}$,
we get $\Gamma_0\!=\!219\pm22\,\mbox{eV}$.
This shows that the theoretical uncertainties of the dispersive calculation are
small.

Unfortunately, the experimental situation is not
clear \cite{PDG}. The value of $\Gamma_{\eta\rightarrow
\pi^+\pi^-\pi^0}$ relies on the rate of the decay into two photons.
The two different methods of measuring $\Gamma_{\eta\rightarrow
\gamma\gamma}$ (photon--photon collisions and
Primakoff effect) yield conflicting results.
While the data based
on the Primakoff effect are in perfect agreement with the
number $Q= 24.2$, which follows from the Dashen theorem,
the $\gamma\gamma$ data
yield a significantly lower result (see l.h.s. of fig. 2).
The statistics is
dominated by the $\gamma\gamma$ data. Using the overall fit of the Particle
Data Group,
$\Gamma_{\eta\rightarrow\pi^+\pi^-\pi^0}\!=\!283\pm28\,\mbox{eV}$ \cite{PDG},
and adding errors quadratically, we obtain $Q\!=\!22.7\pm 0.8$, to be compared
with the result $Q\!=\!22.4\pm0.9$ given in ref. 
\cite{Kambor Wiesendanger Wyler}. With this value of $Q$, the low energy 
theorem (\ref{defQ}) implies that the electromagnetic self energy
amounts to $(M_{K^+}\!-\!M_{K^0})_{e.m.} \!=\! 2$ MeV, to within an
uncertainty of the order of 20\%, in agreement with the lattice result.  
I conclude that, within the
remarkably small errors of the individual determinations,
the two different methods of measuring $Q$ are consistent with each other, but
repeat that one of these relies on the lifetime of the $\eta$, where the
experimental situation is not satisfactory.

8. Kaplan and Manohar \cite{Kaplan Manohar} pointed out that a change in the 
quark masses of the form 
$m_u'\!=\!m_u+\alpha\,
m_d\,m_s$ $(\mbox{cycl}.\;u\!\rightarrow\!d\!\rightarrow\!s\rightarrow\!u$) 
may be absorbed in a change of the 
effective coupling constants 
$L_6,\,L_7,\,L_8$.
The results obtained with the
effective Lagrangian for the meson masses, scattering amplitudes and
matrix elements of the vector and axial currents are invariant under
the operation. Conversely,
since the ratios $m_u/m_d$ and $m_s/m_d$
do not remain invariant, they cannot be
determined with the experimental low energy information concerning these
observables\footnote{The
transformation maps the elliptic constraint onto itself: to first order
in isospin breaking, the quantitiy $1/Q^2$ may equivalently be written as
$(m_d^2-m_u^2)/(m_s^2-m_d^2)$, and the
differences $m_d^2-m_u^2\,,\,m_s^2-m_d^2\,,\,m_u^2-m_s^2$ are invariant.}.
In particular, phenomenology by itself does not exclude the value $m_u\!=\!0$,
widely discussed in the
literature \cite{Banks Nir Seiberg}, as a possible solution of the strong CP 
puzzle.

We are not dealing with a symmetry of QCD, nor is the effective Lagrangian
intrinsically ambiguous: even at the level of the
effective theory, the
predictions for the matrix elements of the scalar and
pseudoscalar operators are not invariant under the above transformation. 
Since an experimental probe sensitive to these is not available, however, 
the size of the correction $\Delta_M$ in eq. (\ref{r1}) cannot be determined 
on purely phenomenological grounds -- theoretical input is needed for this 
purpose.
In the following, I use the $1/N_c$ expansion and the requirement that 
SU(3) represents a
decent approximate symmetry. For a more detailed discussion of the issue,
I refer to \cite{HL Dallas Florida ITEP}.

9. The problem
disappears in the large-$N_c$ limit, because the transformation
$m_u'\!=\!m_u+\alpha\,m_d\,m_s$
violates the Zweig rule \cite{Gerard,HL 1990}. For $N_c\!\rightarrow\!\infty$,
the quark loop graph that gives rise to the U(1) anomaly
is suppressed, so that QCD acquires an additional
symmetry, whose spontaneous breakdown gives rise to a ninth Goldstone
boson, the $\eta'$.
The implications for the effective Lagrangian are extensively discussed in
the literature, and the leading terms in the
expansion in powers of $1/N_c$ were worked out long ago \cite{Leff U(3)}.
More recently, the analysis was extended to first non-leading order, accounting
for all terms which are suppressed either by one power of $1/N_c$ or by one
power of the quark mass matrix \cite{bound}.
This framework leads to the bound
\be \Delta_M>0\co\ee
which excludes the hatched region in fig. 2.
Since the
Weinberg ratios correspond to $\Delta_M\!=\!0$, they
are located at the boundary of this region. In view
of the elliptic constraint, the bound in particular implies
$m_u/m_d\,\raisebox{0.2em}{$>$}\hspace{-0.8em}
\raisebox{-0.3em}{$\sim$}\,\frac{1}{2}$ and thus excludes a massless $u$-quark.
\begin{figure}[t]
\centering
\mbox{\epsfysize=8cm \epsfbox{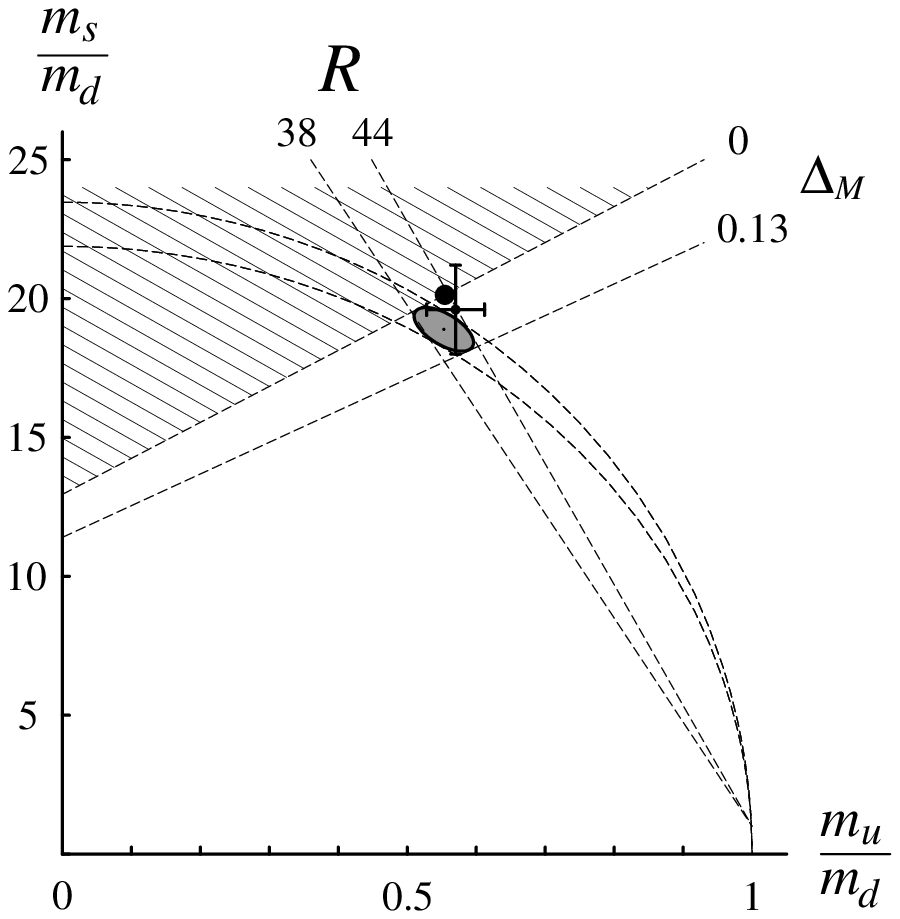} }
\parbox{13.7cm}{Figure 2: Quark mass ratios. The dot corresponds to
Weinberg's
values, while the cross represents the estimates given in ref. \cite{Phys
Rep}. The hatched region is excluded by the bound $\Delta_M>0$. The
error ellipse shown is characterized by the constraints $Q=22.7\pm
0.8$, $\Delta_M>0$, $R<44$, which are indicated by dashed lines.} \end{figure}

10. An upper limit for $m_u/m_d$ may be obtained from the
branching ratio
$\Gamma_{\psi' \rightarrow\psi \pi^0}/\Gamma_{\psi' \rightarrow\psi\eta}$.
Disregarding electromagnetic contributions \cite{DW psi prim},
the ratio of
transition amplitudes is proportional to $(m_d-m_u)/(m_s-\hat{m})$:
\bdm \frac{T_{\psi' \rightarrow\psi+\pi^0}}
          {T_{\psi'\rightarrow\psi+\eta}}
=\frac{3\sqrt{3}}{4\,R}\,(1+\Delta_{\psi'})\co\;\;\;
\frac{1}{R}\equiv\frac{m_d-m_u}{m_s-\hat{m}}\fs\edm
SU(3) predicts that,
for quarks of equal mass, $\Delta_{\psi'}$ vanishes: this term
represents an SU(3)-breaking effect of order $m_s-\hat{m}$.
The data on the branching ratio imply $R\!=\!(31\pm 4)\,(1+\Delta_{\psi'})$,
where the given error only accounts for the experimental accuracy.
The breaking of SU(3) is analysed in ref. \cite{DW psi prim},
on the basis of the multipole expansion. The calculation yields a remarkably
small result for $\Delta_{\psi'}$, indicating a value of $R$ close to 31,
but the validity of 
the multipole expansion for the relevant transition matrix elements is
doubtful \cite{Luty Sundrum}. Moreover, fig. 2 shows that
the result of this calcuation is in conflict with the large-$N_c$ bound.
Since the quark mass ratios given in
refs. \cite{DHW mass ratios} rely on the value of $R$ obtained in this way,
they face the same objections.

At the present level of
theoretical understanding, the magnitude of $\Delta_{\psi'}$ is
too uncertain to allow a determination of $R$, but I do
not see any reason to doubt that SU(3) represents a decent approximate
symmetry also for charmonium. The scale of first order SU(3) breaking effects
such as $\Delta_M$, $(F_K-F_\pi)/F_\pi$ or
$\Delta_{\psi'}$ is set by
$(M_K^2-M_\pi^2)/M_0^2\!\simeq\! 0.25$. Indeed, a correction of this size
would remove the discrepancy with the
large-$N_c$ bound. Large values of $R$, on the other
hand, are inconsistent with the eightfold way. As a conservative upper limit
for the breaking of SU(3), I use $|\Delta_{\psi'}|\!<\!0.4$. Expressed
in terms of $R$, this implies $R<44$. The value $m_s/\hat{m}\!=\!29\pm7
$, obtained by Bijnens, Prades and de Rafael with QCD sum rules
\cite{BPdR}, yields an independent check: the lower end of this interval
corresponds to $\Delta_M< 0.17$. Figure 2 shows that this constraint
also restricts
the allowed region to the right and is only slightly weaker than the
above condition on $R$.

11. The net result for the
quark mass ratios is indicated by the shaded error ellipse in fig. 2, which is
defined by the following three constraints: (i) On the
upper and lower sides, the ellipse is bounded by the two dashed lines
that correspond
to $Q=22.7\pm0.8$. (ii) To the left, it touches the hatched region,
excluded by the large-$N_c$ bound. (iii) On the right, I use the upper limit
$R<44$, which follows from the observed value of the
branching ratio
$\Gamma_{\psi'\rightarrow \psi\pi^0}/
\Gamma_{\psi'\rightarrow \psi\eta}$.
The corresponding range of the various parameters of interest is
\bea \frac{m_u}{m_d}=0.553\pm0.043\co\;\;\;
     \frac{m_s}{m_d}\al=\al18.9\pm0.8\co\;\;\;
     \frac{m_s}{m_u}=34.4\pm3.7\co\no
 \frac{m_s-\hat{m}}{m_d-m_u}= 40.8\pm 3.2\co\;\;\;
\frac{m_s}{\hat{m}}\al=\al 24.4\pm1.5\co\;\;\;
\Delta_M
= 0.065\pm0.065
\fs\nonumber\eea

While
the central value for $m_u/m_d$ happens to coincide with the leading
order formula, the one for $m_s/m_d$ turns out to be slightly smaller. The
difference, which amounts to
6\%, originates in the fact that the
available
data on the $\eta$ lifetime as well as the lattice result for
the
electromagnetic self energies of the kaons imply a somewhat smaller value of
$Q$ than
what is predicted by the Dashen theorem, in agreement with ref. 
\cite{DHW mass ratios}. 

The result for the ratio of isospin- to 
SU(3)-breaking mass differences, $R\!=\!40.8\pm3.2$,  
confirms the early determinations
described in \cite{Phys Rep}. As shown there, the mass
splittings in the baryon octet yield three independent estimates of $R$, i.e.
$51\pm 10\;(N\!-\!P)$, $43\pm 4 \;(\Sigma^-\!-\!\Sigma^+)$ and
$42\pm 6\; (\Xi^-\!-\!\Xi^0)$\footnote{Note that, in this case, the expansion
contains terms of order $m^{\frac{3}{2}}$, which do
play a significant role numerically. The error bars represent simple
rule-of-thumb estimates, indicated by the
noise visible in the calculation. For details see ref. \cite{Phys Rep}.}.
These numbers are perfectly consistent with the
value given above.
A recent
reanalysis of $\rho\!-\!\omega$ mixing \cite{Urech1} leads to
$R\!=41\pm4$ and thus corroborates the picture further. 

I find it remarkable that, despite the problems generated by the 
determinant of the Dirac operator for quark masses of realistic size, 
the lattice results for the mass ratios  
are quite close to the 
above numbers. The most recent values are $m_u/m_d\!=0.512\pm0.006$, 
$(m_d-m_u)/m_s\!=\! 0.0249\pm0.0003$, where the error only accounts for the 
statistical noise \cite{Duncan Eichten Thacker}. They correspond to 
$Q\!=\!22.9$, $\Delta_M\!=\!0$, $R\!=\!38.6$ -- the place where the error 
ellipse shown in fig. 2 touches the large-$N_c$ bound.
 
Finally, I use the value of $m_s$ obtained with QCD sum rules 
\cite{QCD SR,BPdR}
as an input and calculate $m_u$ and $m_d$ with the above ratios. The result 
for the running masses in the $\overline{\mbox{MS}}$ scheme at scale 
$\mu\!=\! 1\,\mbox{GeV}$ reads
\bdm m_u=5.1\pm 0.9\,\mbox{MeV}\co\;\;\;   m_d=9.3\pm\,1.4\,\mbox{MeV}\co\;\;
m_s=175\pm 25 \mbox{MeV}\fs
\edm

\end{document}